%% file: Sage_asplos21.tex
\documentclass[10pt]{jpaper}

\usepackage{etex}
\usepackage[sort,nocompress,adjust]{cite} 

\usepackage{amsmath,amssymb,amsfonts}
\usepackage{algorithmic}
\usepackage{graphicx}
\usepackage{listings}
\usepackage{xcolor}
\definecolor{olive}{rgb}{0.33, 0.42, 0.18}
\usepackage{makecell}
\usepackage{textcomp}
\usepackage{fancyhdr}
\usepackage{tabu}
\usepackage{enumitem}
\usepackage[normalem]{ulem}
\usepackage{inconsolata}
\usepackage[scaled]{beramono}
\usepackage{float}
\usepackage{caption}

\def\BibTeX{{\rm B\kern-.05em{\sc i\kern-.025em b}\kern-.08em
T\kern-.1667em\lower.7ex\hbox{E}\kern-.125emX}}

\usepackage{verbatim}   

\usepackage{wrapfig}
\usepackage{multirow}
\usepackage{balance}
\usepackage{cite}

\usepackage{pifont}

\newcommand{\smallcapital}{\fontsize{9pt}{10pt}\selectfont}

\usepackage{rotating}

\usepackage{arydshln}

\usepackage{placeins}

\usepackage{gensymb}

\usepackage{enumitem}
\usepackage[ruled,vlined]{algorithm2e}
\usepackage{upgreek,textgreek}
\usepackage[htt]{hyphenat}
\usepackage{appendix}

\hypersetup{colorlinks,
	linkcolor=blue,
	citecolor=blue,
	urlcolor=blue,
	filecolor=blue}

\usepackage{mathtools}

\usepackage{tikz}
\newcommand*\circled[1]{\tikz[baseline=(char.base)]{
						\node[shape=circle,draw,inner sep=0.5pt] (char) {#1};}}

\usepackage{listings}
\usepackage[utf8]{inputenc}
\lstdefinestyle{customcpp}{
	 aboveskip=0in,
	  belowskip=0in,
	   abovecaptionskip=0in,
	    belowcaptionskip=0in,
	     captionpos=b,
	      xleftmargin=\parindent,
	       language=C++,
	        morekeywords={forall},
		 showstringspaces=false,
		  basicstyle={\linespread{0.6}\fontseries{sb}\small\ttfamily},
		   keywordstyle=\bfseries,
		    commentstyle=\itshape\color{green!40!black},
	    }

\begin{document}

%
\title{Sage: Using Unsupervised Learning for Scalable Performance Debugging in Microservices}

%
\author{$^1$Yu Gan, $^1$Mingyu Liang, $^2$Sundar Dev, $^2$David Lo, and $^1$Christina Delimitrou\\ $^1$Cornell University, $^2$Google\\Contact author: yg397@cornell.edu}

\date{}

\maketitle

\begin{abstract}
\input{Abstract.tex}
\end{abstract}


%
%
\input{Introduction.tex}

\input{Related_Work.tex}

\input{Techniques.tex}
\input{Design.tex}


\input{Methodology.tex}

\input{Evaluation.tex}

\input{Future_Work_and_Conclusions.tex}


%

\balance
%

\bibliographystyle{IEEEtranS}
\bibliography{references}


\end{document}

%% file: Abstract.tex
Cloud applications are increasingly shifting from large monolithic services 
to complex graphs of loosely-coupled microservices. Despite the advantages of modularity 
and elasticity microservices offer, they also complicate cluster management and performance debugging,  
as dependencies between tiers introduce backpressure and cascading QoS violations. 

We present \textit{Sage}, a machine learning-driven root cause analysis system for 
interactive cloud microservices. Sage leverages unsupervised ML models 
to circumvent the overhead of trace labeling, captures the impact of dependencies between microservices 
to determine the root cause of unpredictable performance online, and applies corrective actions to recover a cloud service's QoS. 
In experiments on both dedicated local clusters and large clusters on Google Compute Engine we show that Sage consistently achieves 
over 93\% accuracy in correctly identifying the root cause of QoS violations, and improves performance predictability. 

\vspace{-0.06in}

%% file: Introduction.tex
\section{Introduction}






Cloud computing has reached proliferation by offering \textit{resource flexibility}, \textit{cost efficiency}, 
and \textit{fast deployment}~\cite{BarrosoBook,tailatscale,Meisner11,Delimitrou13,Delimitrou14,EC2,ContainerEngine}. 
As the scale and complexity of cloud services increased, their design started undergoing a major shift. 

In place of large \textit{monolithic} services that encompassed the entire functionality in a single binary, 
cloud applications have progressively adopted fine-grained modularity, consisting 
of hundreds or thousands of single-purpose and loosely-coupled 
\textit{microservices}~\cite{Cockroft15,Cockroft16,twitter_decomposing,Suresh17,gan:asplos:2019:microservices,Gan18b,gan:asplos:2019:seer,usuite}. 
This shift is increasingly pervasive, with cloud-based services, such as Amazon, Twitter, Netflix, 
and eBay, having already adopted this application model~\cite{Cockroft15,Cockroft16,twitter_decomposing}. 
There are several reasons that make microservices appealing, including the fact that they accelerate and 
facilitate development, they promote elasticity, and enable software heterogeneity, only requiring 
a common API for inter-microservice communication. 

Despite their advantages, microservices also introduce new system challenges. They especially complicate 
resource management, as dependencies between tiers introduce backpressure effects, causing 
unpredictable performance to propagate through the system~\cite{gan:asplos:2019:microservices,gan:asplos:2019:seer}. 
Diagnosing such performance issues empirically is both cumbersome and prone to errors, especially as 
typical microservices deployments include hundreds or thousands of unique tiers. Similarly, current 
cluster managers~\cite{Delimitrou13,Delimitrou14,Delimitrou15,Borg,Lin11,Meisner11,Lo14,Lo15,Mars13a,Mars13b,Nathuji07,Nathuji10,Ousterhout13,omega13,Cloudscale} 
are not expressive enough to account for the impact of microservice dependencies, thus putting more 
pressure on the need for automated root cause analysis systems. 

Machine learning-based approaches have been effective in cluster management 
for batch applications~\cite{resourcecentral}, and for batch and interactive, single-tier services~\cite{Delimitrou13,Delimitrou14}. 
On the performance debugging front, there has been increased attention on trace-based methods to analyze~\cite{xtrace,Ousterhout15,Chen08}, 
diagnose~\cite{Aguilera03,Wang04,Jin12,Reynolds05,Wang11,Cohen04,Teoh19,Nagaraj12,xray,Cherkasova08,Elmroth15,Xie12}, 
and in some cases anticipate~\cite{gan:asplos:2019:seer,Gan18b,Tan12} performance issues in cloud services. 
While most such systems target cloud applications, the only one focusing on microservices is Seer~\cite{gan:asplos:2019:seer}. Seer leverages 
a deep learning model to anticipate upcoming QoS violations, and adjusts the resources per microservice to avoid them. 
Despite its high accuracy, Seer uses supervised learning, which requires offline and online trace labeling, 
as well as considerable kernel-level instrumentation and fine-grained tracing to track the number of outstanding requests across the system stack. 
In a production system this is non-trivial, as it involves injecting resource contention in live applications, which can impact performance and user experience. 






We present Sage, a root cause analysis system that 
leverages unsupervised learning to identify the culprit of unpredictable performance 
in complex graphs of microservices. Specifically, Sage uses Causal Bayesian Networks 
to capture the dependencies between microservices, and counterfactuals through a Graphical 
Variational Autoencoder to examine the impact of microservices on end-to-end performance. 
Sage does not rely on data labeling, hence it can 
be entirely transparent to both cloud users and application developers, scales well 
with the number of microservices and machines, and only relies on lightweight tracing 
that does not require application changes or kernel instrumentation. We have evaluated 
Sage both on dedicated local clusters and large GCE settings with several end-to-end 
microservices~\cite{gan:asplos:2019:microservices}, and showed that it correctly 
identifies the microservice(s) and system resources that initiated a QoS violation in over 93\% of cases, 
and improves performance predictability without sacrificing efficiency. 

\vspace{-0.02in}


%% file: Related_Work.tex
\vspace{-0.04in}
\section{Related Work}
\label{sec:relatedWork}

Below we review work on the system implications of microservices, cluster managers 
designed for multi-tier services and microservices, and systems for cloud performance debugging. 

\vspace{-0.06in}
\subsection{System Implications of Microservices}
\vspace{-0.08in}

The increasing popularity of fine-grained modular application design, microservices being an extreme
materialization of it, has yielded a large amount of prior work on
representative benchmark suites and studies on their characteristics~\cite{sirius,usuite,gan:asplos:2019:microservices}.
$\mu$Suite~\cite{usuite} is an open-source
multi-tier application benchmark suite containing several online data-intensive (OLDI) services,
such as image similarity search, key-value stores, set intersections, and recommendation systems.
DeathStarBench~\cite{gan:asplos:2019:microservices} presents five end-to-end interactive applications
built with microservices, leveraging Apache Thrift~\cite{thrift}, Spring Framework~\cite{spring}, and gRPC~\cite{grpc}.
The services implement popular cloud applications, like social networks, e-commerce sites, and movie reviewing services.
DeathStarBench also explores the hardware/software implications of microservices, including their
resource bottlenecks, OS/networking overheads, cluster management challenges, and sensitivity to performance unpredictability.
Accelerometer~\cite{accelerometer} characterizes the system overheads of several Facebook microservices,
including I/O processing, logging, and compression. They also build an analytical model
to predict the potential speedup of a microservice from hardware acceleration.


\vspace{-0.06in}
\subsection{Microservices Cluster Management}
\vspace{-0.08in}

Microservices have complicated dependency graphs, strict QoS targets, and are sensitive to performance unpredictability.
Recent work has started exploring the resource management challenges of microservices.
Suresh et al.~\cite{Suresh17} design Wisp, a dynamic rate limiting system for microservices, which
prioritizes requests in the order of their deadline expiration. 
uTune~\cite{utune} auto-tunes
the threading model of multi-tier applications to improve their end-to-end performance.
GrandSLAm~\cite{GrandSLAm} improves the resources utilization of ML microservices by estimating
the execution time of each tier, and dynamically batching and reordering requests to meet QoS.
Finally, SoftSKU~\cite{softsku} characterizes the performance of the same Facebook microservices as~\cite{accelerometer}
across hardware and software configurations, and searches for their optimal resource configurations 
using A/B testing in production.


\vspace{-0.06in}
\subsection{Cloud Performance Debugging}
\vspace{-0.08in}

There is extensive prior work on monitoring and debugging performance and efficiency issues in cloud systems.
Aguilera et al.~\cite{Aguilera03} built a tool to construct the casual path of a service from RPC messages without access to source code. 
X-Trace~\cite{xtrace} is a tracing framework portable across protocols and software systems that detects
runtime performance issues in distributed systems. It can identify faults in several scenarios, including DNS resolution
and overlay networks. Mystery Machine~\cite{MysteryMachine} leverages a large amount
of cloud traces to infer the causal relationships between requests at runtime.
There are also several production-level distributed tracing systems, including Dapper~\cite{dapper}, Zipkin~\cite{zipkin},
Jaeger\cite{jaeger}, and Google-Wide Profiling (GWP)~\cite{gwp}. Dapper, Zipkin and Jaeger
record RPC-level traces for sampled requests across the calling stack, while GWP monitors low-level hardware metrics.
These systems aim to facilitate locating performance issues, but are not geared towards taking action to resolve them. 

Autopilot~\cite{eurosys20autopilot} instead adjusts the number of tasks and CPU/memory limits automatically
to reduce resource slack while guaranteeing performance. 
Sage differs from prior work on cloud scheduling, such as~\cite{icac14autoscaling,Mars11a,Mars13a,Delimitrou14}, in that 
it locates the root cause of poor performance only using the end-to-end QoS target, without explicitly requiring to define per-tier performance service level agreements (SLAs). 

Root cause analysis systems for cloud applications are gaining increased attention, 
as the number of interactive applications continues to increase. Several of these proposals
leverage statistical models to diagnose performance issues~\cite{Wang11,Xie12,Tan12}. 
Cohen et al.~\cite{Cohen04} build tree-augmented Bayesian networks (TANs) to predict whether QoS
will be violated, based on the correlation between performance and low-level metrics. Unfortunately, in multi-tier
applications, correlation does not always imply causation, given the existence of backpressure effects between dependent tiers.
ExplainIt!~\cite{ExplainIt} leverages a linear regression model to find
root causes of poor performance in multi-stage data processing pipelines which optimize for throughput. 
While the regression model works well for batch jobs, latency is more sensitive to noise, and propagates across dependent tiers.

CauseInfer~\cite{causeinfer} as well as Microscope~\cite{lin2018microscope} build a causality graph
using the PC-algorithm, and use it to identify root causes with different anomaly detection algorithms. 
As with ExplainIt!, they work well for data analytics, but would be impractical for latency-critical applications with tens
of tiers, due to the high computation complexity of the PC-algorithm~\cite{kalisch2007estimating}.
Finally, Seer~\cite{gan:asplos:2019:seer} is a supervised CNN+LSTM model that anticipates QoS violations
shortly before they happen. Because it is proactive, Seer can avoid poor performance altogether,
however, it requires considerable kernel-level instrumentation to track the number of outstanding requests
across the system stack at fine-granularity, which is not practical in large production systems. It also requires
data labeling to train its model, which requires injecting QoS violations in active services.
This sensitivity to tracing frequency also exists in Sieve~\cite{sieve-middleware}, which uses the Granger
causality test to determine causal relationships between tiers~\cite{silvestrini2008temporal,anderson2019sensitivity}.



%% file: Techniques.tex
\section{ML for Performance Debugging}
\label{section:techniques}

\vspace{-0.06in}
\subsection{Overview}
\vspace{-0.06in}

\begin{figure}[tb]
  \centering
  \includegraphics[width=0.45\textwidth]{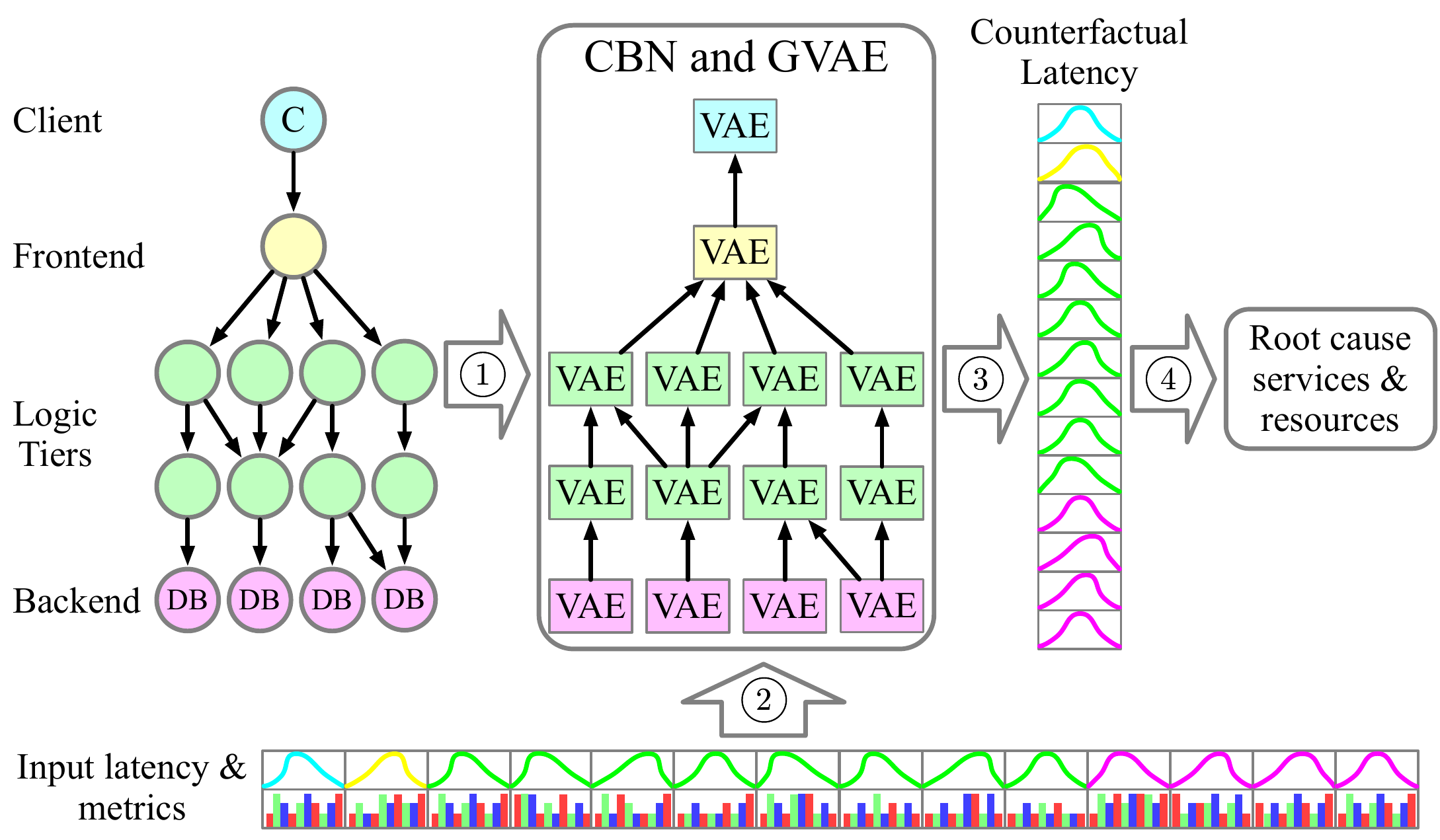}
  \vspace{-0.05in}
  \caption{Sage's ML pipeline.
    \protect\circled{1}: Build CBN and GVAE.
    \protect\circled{2}: Process per-tier latency and usage.
    \protect\circled{3}: Generate counterfactuals with GVAE.
    \protect\circled{4}: Identify root cause services \& resources. }
  \label{fig:ml_pipeline}
  \vspace{-0.10in}
\end{figure}

Sage is a performance debugging and root cause analysis system for large-scale cloud applications. While the design centers around interactive microservices,
where dependencies between tiers further complicate debugging, Sage is also applicable to monolithic architectures. 
Sage diagnoses the \textit{root cause}~\cite{sep-causation-probabilistic} of end-to-end QoS violations, and applies appropriate corrective action to restore performance.
Fig.~\ref{fig:ml_pipeline} shows an overview of Sage's ML pipeline.
Sage relies on two techniques, each of which is described in detail below; first, it automatically captures the dependencies between microservices
using a Causal Bayesian Network ({\smallcapital CBN}) trained on RPC-level distributed traces~\cite{dapper,zipkin,gan:asplos:2019:seer,Gan18b}. 
The CBN also captures the latency propagation from the backend to the frontend. 
Second, Sage uses a graphical variational auto-encoder ({\smallcapital GVAE}) to generate hypothetical scenarios (counterfactuals~\cite{sep-causation-law}),
which tweak the performance and/or usage of individual microservices to values known to meet QoS, and infers whether the change restores QoS. 
Using these two techniques, Sage determines which set of microservices initiated a QoS violation,
and adjusts their deployment or resource allocation. 

While prior work has highlighted the potential of ML for cloud performance debugging~\cite{gan:asplos:2019:seer},
such techniques rely exclusively on supervised models, which require injecting resource contention on active services
to correctly label the training dataset with root causes of QoS violations~\cite{gan:asplos:2019:seer}.
This is problematic in practice, as it disrupts the performance of live services. 
Additionally, prior work requires high tracing frequency and heavy instrumentation to collect metrics like the number of outstanding
requests across the system stack, which is not practical in a production system and can degrade performance. 

Sage instead adheres to the following design principles:
\begin{itemize}[leftmargin=*]
  \item \textbf{Unsupervised learning}: Sage 
        does not require labeling training data, and it diagnoses QoS violations using low-frequency traces collected during live traffic
        using tracing systems readily available in most major cloud providers.
  \item \textbf{Robustness to sampling frequency}: Sage does not require tracking individual requests to detect temporal patterns,
        making it robust to tracing frequency. This is important, as production tracing systems like Dapper~\cite{dapper} employ
        aggressive sampling to reduce overheads~\cite{google-cloud-monitoring, aws-cloudwatch}.
        In comparison, previous studies~\cite{gan:asplos:2019:seer, sieve-middleware, epsilon-diagnosis-www} collect traces
        at millisecond granularity, which can introduce significant overheads.
  \item \textbf{User-level metrics}: Sage only uses user-level metrics, easily obtained
        through cloud monitoring APIs and service-level traces from distributed tracing frameworks,
        such as Jaeger~\cite{jaeger}. It does not require any kernel-level information,
        which is expensive, or even inaccessible in cloud platforms.
  \item \textbf{Partial retraining}: A major premise of microservices is enabling frequent updates.
        Retraining the entire system every time the code or deployment of a microservice changes is prohibitively expensive.
        Instead Sage implements partial and incremental retraining, whereby only the microservice that changed and its immediate
        neighbors are retrained. 
  \item \textbf{Fast resolution}: Empirically examining sources of poor performance is costly in time
        and resources, especially given the ingest delay cloud systems have in consuming monitoring
        data, causing a change to take time before propagating on recorded traces. Sage models the impact
        of the different probable root causes concurrently, restoring QoS faster. 
\end{itemize}


\subsection{Microservice Latency Propagation}
\label{sec:latency_observations}

\vspace{-0.06in}
\subsubsection{Single RPC Latency Decomposition}
\label{sec:single_rpc}

\begin{figure}
  \centering
  \includegraphics[width=0.46\textwidth]{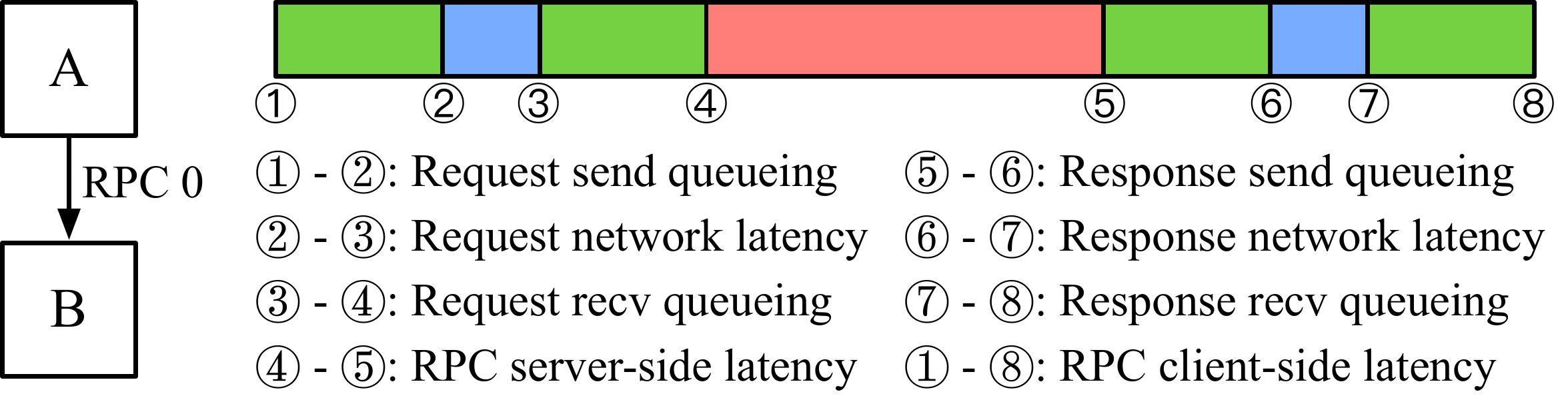}
  \vspace{-0.12in}
  \caption{RPC latency breakdown. Red: RPC client-side latency,
    blue: network latency, green: application queueing. }
  \label{fig:single_rpc}
  \vspace{-0.08in}
\end{figure}

Fig.~\ref{fig:single_rpc} shows the latency decomposition of an RPC across client (sender) and server (receiver).
The client initiates an RPC request via the \texttt{rpc0\_request} API at \circled{1}.
The request then waits in the RPC channel's send queue and gets written to the Linux network stack via the
\texttt{sendmsg} syscall at \circled{2}. The packets pass through the TCP/IP
protocol and are sent out from the client's NIC. They are then transmitted over the wire and switches
and arrive at the server's NIC. After being processed by the server's network protocol stack at \circled{3},
the request is queued in the RPC channel's receive queue, waiting to be processed via the \texttt{rpc0\_handler},
which starts at \circled{4} and ends at \circled{5}.
Finally, the RPC response follows the same process from server to client, until
it is received by the client's application layer at \circled{8}.
\circled{1} - \circled{8} and \circled{4} - \circled{5} are the application-level client-
and server-side latencies, respectively. 
\circled{2} - \circled{3}  and \circled{6} - \circled{7} are the latencies in the network protocol, switches, and wiring.
\circled{1} - \circled{2}, \circled{3} - \circled{4}, \circled{5} - \circled{6},
and \circled{7} - \circled{8} is the queueing time in the application layer of the client and server.

Timestamps for the user-level events \circled{1}, \circled{4}, \circled{5}, and \circled{8} can be obtained
with distributed tracing frameworks, such as Jaeger. Timestamping \circled{2}, \circled{3}, \circled{6}, and \circled{7}, 
would require probing the Linux kernel with high-overhead tools, like SystemTap~\cite{systemtap}.
Instead, we approximate the request/response network delay by measuring the zero-load latency between client and server,
when queueing in the application is zero.

\begin{figure}[htb]
  \centering
  \vspace{-0.12in}
  \includegraphics[width=0.44\textwidth]{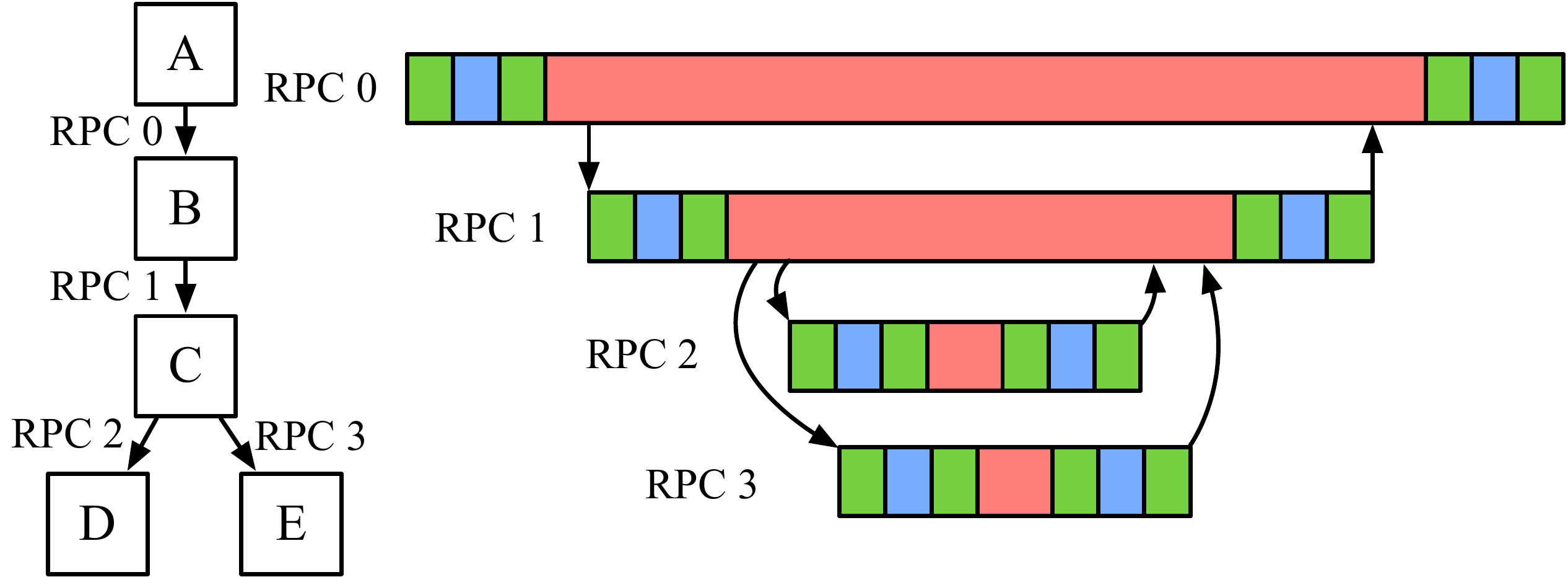}
  \vspace{-0.08in}
  \caption{Dependency graph and traces of nested RPCs. }
  \label{fig:multiple_rpcs}
  \vspace{-0.10in}
\end{figure}

\subsubsection{Markov Property of RPC Latency Propagation}
\label{sec:multiple_rpcs}

Multiple RPCs form a tree of nested traces in a distributed monitoring system. Fig.~\ref{fig:multiple_rpcs} shows
an example RPC dependency graph with five services, four RPCs, and its corresponding latency traces.
When the user request arrives at $A$, it sends \texttt{RPC0} to service $B$. $B$ further forwards the request to $C$ via \texttt{RPC1},
and $C$ sends it to the backend services $D$ and $E$ via \texttt{RPC2} and \texttt{RPC3} in parallel. 
After processing the responses from $D$ and $E$, $C$ replies to $B$, and $B$ replies to $A$, as \texttt{RPC1} and \texttt{RPC0} return.

The server-side latency of any non-leaf RPC is determined by the processing time of the RPC itself
and the queueing time (i.e., client-side latency) of its child RPCs. This latency propagates
through the RPC graph to the frontend. 
Since the latency of a child RPC cannot propagate to its parent without impacting its own latency,
the latency propagation follows a \textit{local Markov property}, where each latency 
is conditionally independent on its non-descendant RPCs, given its child RPC latencies~\cite{koller2009probabilistic}.
For instance, the latency of \texttt{RPC0} is conditionally independent on \texttt{RPC2}
and \texttt{RPC3}, given the latency of \texttt{RPC1}.
\begin{wrapfigure}[10]{l}{0.196\textwidth}
  \centering
  \vspace{-0.24in}
  \hspace*{-0.04in}\includegraphics[width=0.25\textwidth]{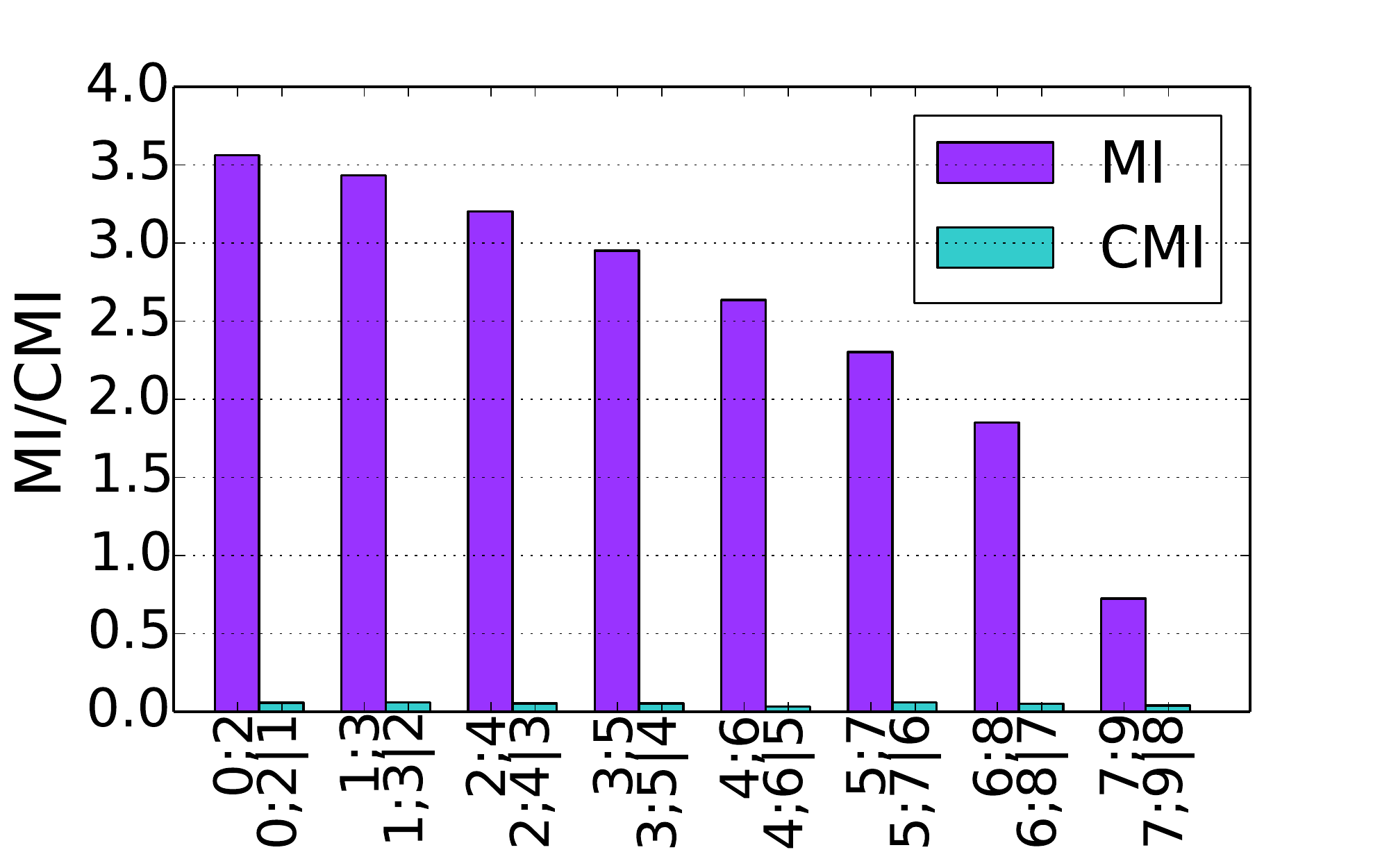}
  \vspace{-0.26in}
  \caption{MI of two distance-of-2 RPCs,
    and CMI given the server latency of the middle RPC.}
  \label{fig:mutual_info}
  \vspace{-0.14in}
\end{wrapfigure}
In information theory, \textit{mutual information} measures the reduction of uncertainty in
one random variable given another random variable. Two random variables are independent
or conditionally independent if their mutual information (MI) or conditional mutual information (CMI) is zero~\cite{gray2011entropy}.
Fig.~\ref{fig:mutual_info} shows the MI of the server-side latencies of two RPCs
with distance of two, and their CMI, given the server-side latency of the
in-between RPC, in a 10-microservice chain. The MI of each two non-adjacent RPCs 
is blocked by the latency of the RPC in the middle, making them conditionally independent~\cite{campos2006scoring}.

\vspace{-0.06in}
\subsection{Modeling Microservice Dependency Graphs}
\vspace{-0.06in}

\subsubsection{Causal Bayesian Networks}

A CBN is a directed acyclic graph (DAG), where the nodes are random variables
and the edges indicate their conditional dependencies, from cause to effect~\cite{neapolitan2004learning,pearl2009causal}.
Sage uses three node types: \vspace{-0.04in}
\begin{itemize}[leftmargin=*]
  \item \textbf{Metric nodes ($X$)}: They contain resource-related metrics of all services and network channels
        collected with tools, like Google Wide Profiling~\cite{google-cloud-monitoring, aws-cloudwatch, azure-monitoring}.
        They are the exogenous variables that cause latency variances across RPCs, and fall into two groups: 
        \textit{server-} and \textit{network-related}. Server-related metrics ($X^s$), include CPU utilization,
        memory bandwidth, context switches, etc., and impact the server's processing time.
        Network-related metrics ($X^{net}$), such as the round trip time (RTT), packet loss rate, network bandwidth, etc.,
        affect the delay of RPC channels.
        The set of necessary and sufficient metrics
        was derived via feature selection. 
        To keep the shape of the vector for each metric the same regardless of
        the replicas per tier, we use a vector of percentiles, e.g., [10th\%, ..., 90th\%, 100th\%]
        averaged across the tier's replicas.

  \item \textbf{Latency nodes ($Y$)}: These include client-side latency ($Y^c$), server-side latency ($Y^s$),
        and request/response network delay ($Y^{req}$ and $Y^{resp}$) of all RPCs of Sec.~\ref{sec:single_rpc}.
        We use a vector of percentiles~\cite{EmpiricalQuantile} to represent the RPC latency distribution.
        Since the RPC tail latency correlates more closely with QoS, 
        high percentiles are sampled more finely. 

  \item \textbf{Latent variables ($Z$)}: These nodes contain the unobservable factors
        that are responsible for latency stochasticity.
        They are critical to generate the counterfactual latencies Sage relies on to diagnose root causes (Sec.~\ref{sec:gvae}).
        We divide latent variables to server-related variables ($Z^s$) which capture individual microservices,
        and network-related variables ($Z^{net}$), which capture links between them. 
\end{itemize}
\vspace{-0.04in}

\begin{wrapfigure}[13]{l}{0.25\textwidth}
  \centering
  \includegraphics[scale=0.31,viewport= 0 50 400 370]{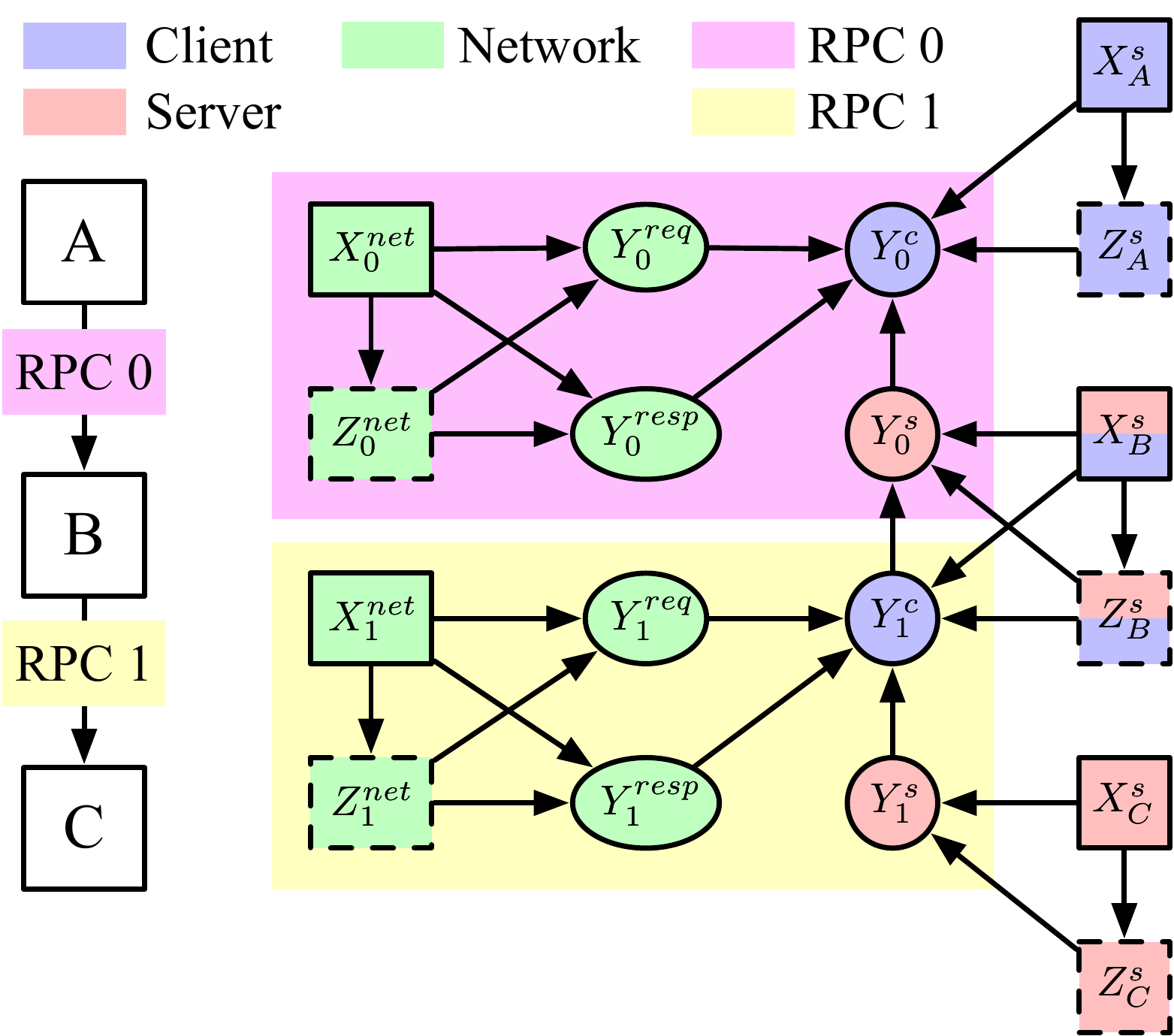}
  \caption{The RPC graph of a 3-service chain, and its corresponding CBN. }
  \label{fig:cbn_two_rpcs}
\end{wrapfigure}

We then construct the CBN among the three node
classes for all RPCs, based on their causal relationships and latency propagation 
obtained via the distributed tracing system (Sec.~\ref{sec:latency_observations}).
Note that metric nodes in the CBN have no causes because they are exogenous variables set outside the model.
Since the distribution of a latent variable is modulated by its corresponding metric node, there is an edge from $X$ to $Z$.
Figure~\ref{fig:cbn_two_rpcs} shows an example of the CBN of a three-microservice dependency chain.
The nodes with solid lines ($X$ and $Y$) are observed, while the nodes with dashed lines ($Z$) are latent variables
that need to be inferred. The arrows in the RPC graph and CBN have opposite directions
because the latency of one RPC is determined by the latency of its child RPCs.

\vspace{-0.08in}
\subsubsection{Latency Distribution Factorization}

We consider the microservice latencies and usage metrics in the CBN to be random and i.i.d variables
from the underlying distribution. 
Using the CBN, we can factorize the joint distribution
into the product of individual tier distributions, conditional on their parent variables. Factorization is
needed to later build the graphical model of Sec.~\ref{sec:gvae}, which will explore
possible root causes. 
We are interested in the following distributions:

\begin{itemize}[leftmargin=*]
  \item The conditional distribution of latency given the observed metrics and latent variables $P(Y \mid X, Z)$,
  \item The prior distribution of latent variables $Z$ given the observed metrics, $P(Z \mid X)$, and
  \item The posterior distribution of latent variables $Z$, given the observed metrics and latency values $Q(Z \mid X, Y)$.
\end{itemize}


\vspace{-0.08in}
\subsection{Counterfactual Queries}
\label{sec:counterfactual_queries}
\vspace{-0.06in}

Sage uses counterfactual queries~\cite{pearl2009causal, morgan2015counterfactuals}
to diagnose the root cause of unpredictable performance. 
In a typical cloud environment, site reliability engineers (SREs) can verify if a suspected root cause is correct
by reverting a microservice's version or resource configuration to a state known to be safe, while keeping all other factors unchanged, 
and verifying whether QoS is restored. 
Sage uses a similar process, where ``suspected root causes''
are generated using counterfactual queries, which determine causality by asking what the outcome would be
if the state of a microservice had been different~\cite{morgan2015counterfactuals, menzies2008counterfactual, hofler2005causal}.
Such counterfactuals can be generated by adjusting problematic microservices 
in the system in a similar way to how SREs take action to resolve
a QoS violation. The disadvantage of this is that interventions take time, and incorrect root cause assumptions hurt performance and resource efficiency.
This is especially cumbersome when scaling microservices out, spawning new instances, or migrating existing ones. 

Instead, Sage leverages historical tracing data to generate realistic counterfactuals.
There are two challenges in this. 
First, the exact situation that is causing the QoS violation now may not have occurred in the past.
Second, the model needs to account for the latent variables $Z$ which also contribute to the distribution of $Y$.
Therefore, we use a generative model to learn the latent distribution $P(Z \mid X)$ and the latency distribution $P(Y \mid X,Z)$,
and use them to generate counterfactual latencies $Y$, given input metrics $X$. We then use the counterfactuals
to conduct ``but-for'' tests for each service and resource, and discover their causal relationship with the QoS violation.
If, after intervening, the probability of meeting QoS
exceeds a threshold, the intervened metrics caused the violation. 

\vspace{-0.08in}
\subsection{Generating Counterfactuals}
\label{sec:gvae}
\vspace{-0.06in}

Conditional deep generative models, such as the conditional variational autoencoders (CVAE)~\cite{NIPS2015_5775}
and conditional generative adversarial nets~\cite{mirza2014conditional},
are common tools to generate new data
from an original distribution. Generally, they compress a high-dimensional target ($Y$) and tag ($X$)
into low-dimensional latent space variables ($Z$), and use them to generate new data.
Recent studies have showed that these techniques can also be used to generate counterfactuals for causal inference~\cite{louizos2017causal,yoon2018ganite}.

\begin{wrapfigure}[9]{r}{0.235\textwidth}
  \centering
  \includegraphics[scale=0.304,viewport=80 20 400 242]{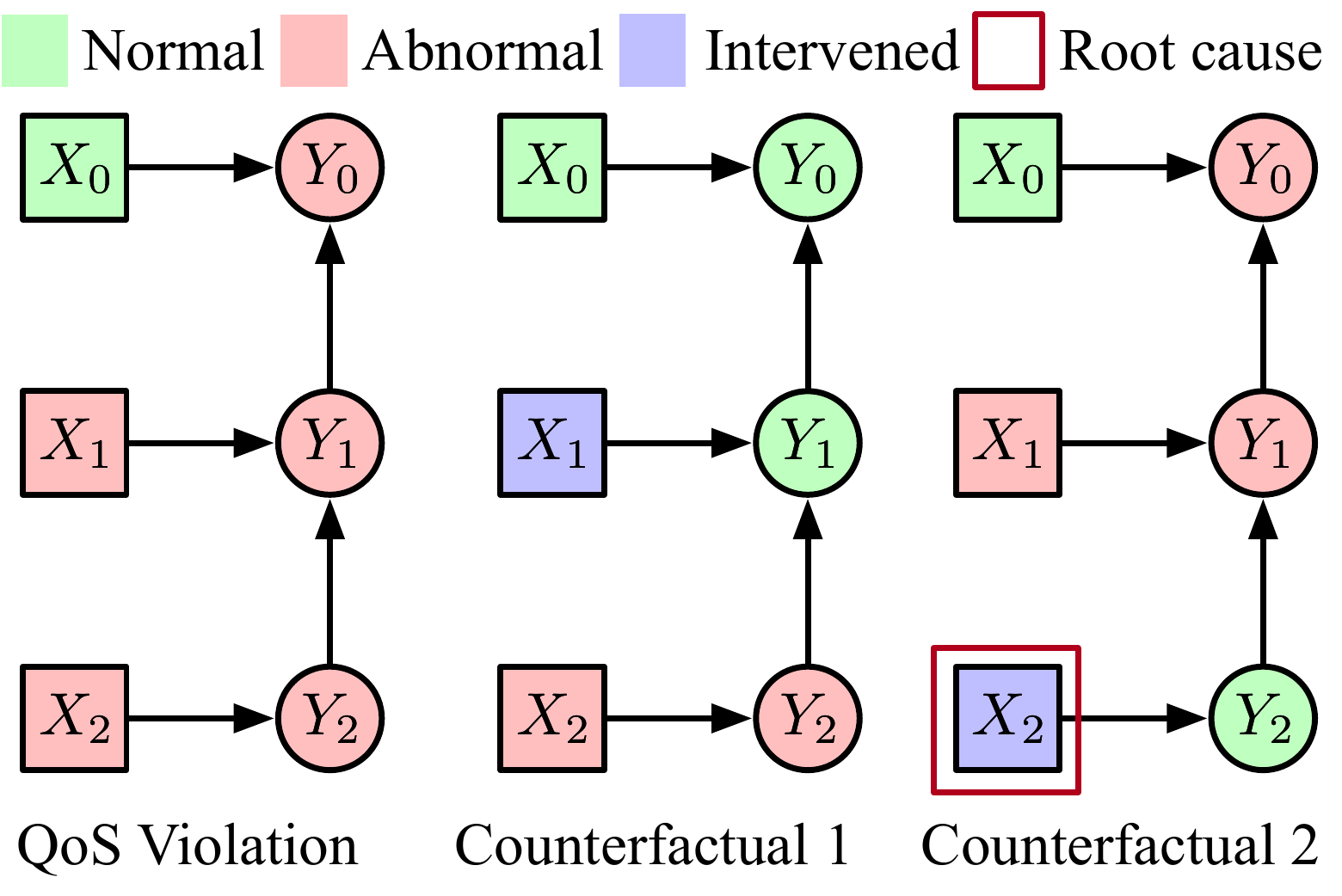}
  \caption{Detecting root causes using counterfactuals. }
  \label{fig:causal_example}
\end{wrapfigure}

Fig.~\ref{fig:causal_example} shows an example
of detecting the root cause of a QoS violation in the 3-tier chain of Fig.~\ref{fig:cbn_two_rpcs}.
Assume that the CPU utilization of Services 1 and 2 is abnormal (different from values that meet QoS).
We evaluate the hypothetical end-to-end latency of two counterfactuals; one where Service 1's utilization is normal,
with all other metrics unchanged, and one where Service 2's utilization is normal.
If fixing Service 1 does not restore QoS, as in \textit{Counterfactual 1}, then
Service 1 \textit{alone} is not the root cause. If fixing the utilization of Service 2 restores QoS, 
as in \textit{Counterfactual 2}, then it is the root cause. Not being enough to restore QoS does not mean
that a service is not part of the problem; if single microservices do not restore 
QoS, Sage considers mixes of tiers. 

To generate counterfactuals, we build a network of CVAEs according to the structure of the CBN.
We adapt the CVAE in~\cite{NIPS2015_5775}, a widely-used hybrid model with a CVAE and a Gaussian stochastic neural network (GSNN).
The CVAE network can be further decomposed into an encoder, decoder, and prior network. During the training phase,
the CVAE receives a mini-batch of $X$ and $Y$ from the training set. The encoder learns
the posterior distribution of $Z$, given the observed $X$ and $Y$, ($Q_{\phi}(Z \mid X, Y)$),
and the prior network learns the prior distribution of $Z$, observing only $X$ ($P_{\psi}(Z \mid X)$).
The decoder then reconstructs the input target $Y$, based on $Z$ sampled from the posterior distribution
and $X$, i.e., $P_\theta(Y \mid X, Z)$, $Z \sim Q_{\phi}(Z \mid X, Y)$. The encoder, decoder, and prior networks
are constructed with multi-layer perceptrons (MLPs) parameterized with $\theta$, $\phi$, and $\psi$, respectively.
During the generation phase, we use the prior network to modulate the distribution of $Z$, given $X$, and use $Z$
sampled from that distribution together with $X$ to generate $Y$. During training, we minimize
the latency reconstruction loss, including a regularization term of Kullback-Leibler (KL) divergence, 
i.e., the negative variational lower bound~\cite{kingma2013auto}:\vspace{-0.08in}
\begin{equation*}
  \label{eq:cvae_cvae}
  \begin{split}
    & L_{\text{CVAE}}(X, Y, Z; \theta, \phi, \psi) = \underbrace{-\mathbb{E}_{Z \sim Q_{\phi}(Z \mid X, Y)} \big[ \log P_{\theta}(Y \mid X, Z) \big]}_{\text{reconstruction loss}} \\
    & + \beta \cdot \underbrace{D_{KL} \big (Q_{\phi}(Z \mid X, Y) \parallel P_\psi(Z \mid X) \big)}_{\text{KL divergence regularization}} \;\;\;\;\;\;\;\;\;\;\;\;\;\;\;\;\;\;\;\; (1)
    \vspace*{-0.28in}
  \end{split}
  \vspace*{-0.18in}
\end{equation*}
\vspace*{-0.04in}
where $\beta > 0$ is the hyperparameter that identifies disentangled latent factors in $Z$~\cite{DBLP:conf/iclr/HigginsMPBGBML17}.
The reconstruction loss term allows the encoder to extract useful input features, and
the decoder to accurately reconstruct the original data from the latent variables. The KL divergence
regularization minimizes overfitting. We further add a GSNN, where $Q_\phi(Z \mid X, Y)=P_\psi(Z \mid X)$, to reconstruct $Y$ by sampling $Z$ from the prior distribution,
to tackle concerns that the CVAE alone may not be enough to train a conditional generative model, because it uses the encoder's posterior distribution during training
and the prior distribution to draw $Z$ samples during generation~\cite{NIPS2015_5775, ivanov2018variational}.
%

Although using a single CVAE for the entire microservice graph would be simple, it has several drawbacks.
First, it lacks the CBN's structural information which is necessary to avoid ineffectual counterfactuals based on spurious correlations.
Second, it prohibits partial retraining, which is essential for frequently-updated microservices.
Finally, it is less explainable since it does not reveal how the latency of a problematic service propagates to the frontend.
Therefore, we construct one small CVAE per microservice with few fully connected and dropout layers,
and connect the different CVAEs according to the structure of the CBN to form the graphical variational autoencoder (GVAE).
The final loss function is: 
\vspace{-0.28in}

\setcounter{equation}{1}
\begin{equation}
  L_{\text{GVAE}}(X, Y, Z; \theta, \phi, \psi) = \sum_{i=1}^m \Big[ \alpha L_{\text{CVAE}_i} + (1 - \alpha) L_{\text{GSNN}_i}\Big]
  \vspace{-0.18in}
\end{equation}

\noindent where $\text{CVAE}_i$ and $\text{GSNN}_i$ is the CVAE and the \\\noindent Gaussian stochastic network for service $i$. 
The encoders and prior networks are trained entirely in parallel. 
The decoders require the outputs of the parent decoders
in the CBN as inputs, and are trained serially. The maximum depth of the CBN determines the max number of serially-cascaded decoders.

\vspace*{-0.04in}

%% file: Design.tex
\section{Sage Design}
\label{sec:design}

\vspace{-0.06in}
Sage is a root cause analysis system for interactive microservices.
Sage relies on RPC-level tracing to compose a CBN with the microservice topology, 
and per-node tracing for per-tier latency distributions.
Below we discuss Sage's monitoring system (Sec.~\ref{sec:tracing}), training and inference pipeline (Sec.~\ref{sec:root_cause}), its actuator
once a root cause has been identified (Sec.~\ref{sec:design_actuation}),
and how Sage handles application changes (Sec.~\ref{sec:retraining}).

\begin{figure}
  \centering
  \includegraphics[width=0.37\textwidth]{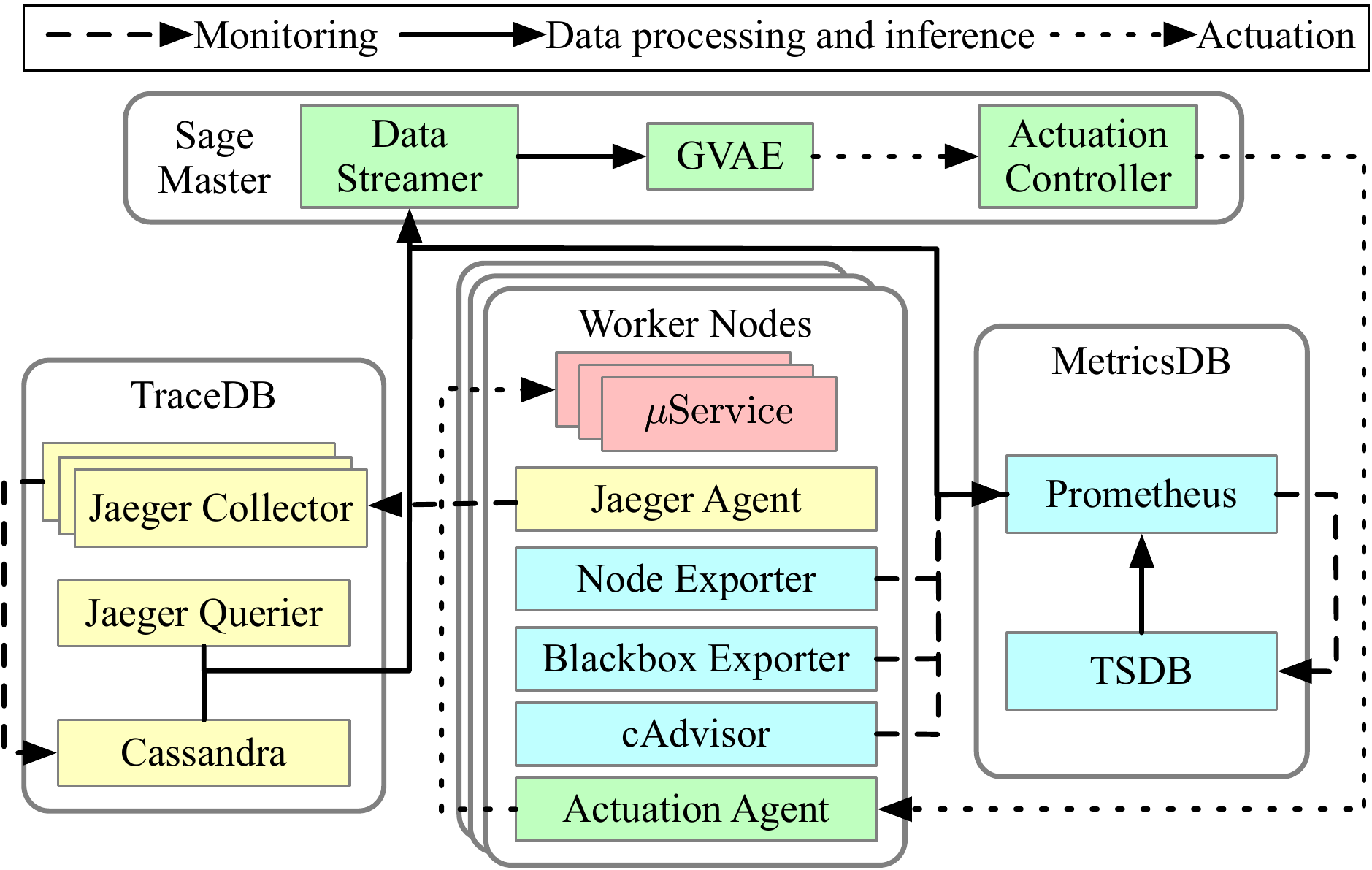}
  \vspace{-0.08in}
  \caption{{Overview of Sage's system design.} }
  \vspace{-0.02in}
  \label{fig:sage_overview}
\end{figure}

Fig.~\ref{fig:sage_overview} shows an overview of Sage. The system uses Jaeger~\cite{jaeger}, a distributed RPC 
tracing system for end-to-end execution traces, and the Prometheus Node Exporter~\cite{nodeexporter}, Blackbox
Exporter~\cite{blackboxexporter}, and cAdvisor~\cite{cadvisor} to collect hardware/OS metrics,
container-level performance metrics, and network latencies. Each metric's timeseries is stored in the
Prometheus TSDB~\cite{prometheus,gorilla}. At runtime, Sage queries Jaeger
and Prometheus to obtain real-time data. The GVAE then 
infers the root cause of any QoS violation(s), at which point 
Sage's actuator adjusts the offending microservice's resources. 

Sage uses a centralized master for trace processing,
root cause analysis, and actuation, implemented in approximately 6KLOC of Python, and per-node agents for trace collection and container deployment.
It also maintains two hot stand-by copies of the master for fault tolerance. The GVAE model is built in PyTorch,
with each VAE's encoder, decoder, and prior network using a DNN with 3-5 fully connected layers,
depending on the input node number. We also use batch normalization between every two hidden layers
for faster convergence, and a dropout layer after the last hidden layer to mitigate overfitting. 


\vspace{-0.06in}
\subsection{Tracing Systems}
\label{sec:tracing}
\vspace{-0.06in}

Sage includes RPC-level latency tracing and container/node-level usage monitoring.
The RPC tracing system is based on Jaeger~\cite{jaeger}, an open-source framework,
similar to Dapper~\cite{dapper} and Zipkin~\cite{zipkin}, and augmented with the Opentracing client library~\cite{opentracing},
to add microservice spans and inject span context to each RPC. It measures each RPC's client- and
server-side latency, and the network latency of each request and response.
Sage records two spans per RPC; one starts when the client sends the RPC request and ends when it receives the response,
while the other starts when the server receives the RPC request and ends when it sends the response to the client, both at application level.
To avoid instrumenting the kernel to measure network latency (Sec.~\ref{sec:single_rpc}),
we use a set of probing requests to calculate the zero-load latency, and infer the request/response
network latency.
We deploy one Jaeger agent per node to retrieve spans for resident microservices.
The Jaeger agents flush the spans to a replicated Jaeger collector for aggregation, which stores
them in a Cassandra database. We additionally enable sampling to reduce tracing overheads, and verify that
with 1\% sampling frequency, the tracing overhead is approximately 2.6\% on the 99th percentile latency
and 0.66\% on the max throughput under QoS. We also ensure that sampling does not lower Sage's accuracy.
Removing sampling would reduce the training time, but would also incur higher overheads~\cite{dapper}.
To account for fluctuations in load, Sage adjusts the sampling and inference
frequency to keep its detection accuracy above a configurable threshold, without incurring high overheads. 

The per-node performance and usage metrics are collected using Prometheus, a widely-used open-source 
monitoring platform~\cite{prometheus}. 
More specifically, we deploy node, blackbox, and cAdvisor exporters
per node to measure the hardware/system metrics, network latency, and container resource usage respectively.
Each metric's timeseries is stored in a centralized Prometheus TSDB. The overhead
of Prometheus is negligible for all studied applications when scraping metrics every 10s. 



\vspace{-0.06in}
\subsection{Root Cause Analysis}
\label{sec:root_cause}
\vspace{-0.06in}

To diagnose a root cause, Sage first relies on the Data Streamer to fetch
and pre-process the tracing data. 
The Streamer queries Jaeger and Prometheus
for an interval's log data over HTTP, and pre-processes them using feature encoding,
aggregation, dimensionality reduction, and normalization.
It outputs RPC latency percentiles across the sampled requests, and performance/usage percentiles
across the replicas of each tier.

Sage initializes and trains the GVAE model offline with all initially-available latency and usage data.
It then periodically retrains the model 
as new requests come in~\cite{chen2018lifelong,parisi2019continual,hoi2018online,yoon2018lifelong}.
Retraining happens even when there are no application changes, 
to account for changes in user behavior. Sage 
handles design changes with partial and incremental retraining to minimize
overheads and accelerate convergence (Sec.~\ref{sec:retraining}).
Every time training is triggered, the GVAE streams in batches of tracing tensors to update its network parameters.
Online learning models are prone to \textit{catastrophic forgetting}, where the model forgets previous knowledge upon
learning new information~\cite{parisi2019continual,kirkpatrick2017overcoming}. To avoid this,
we interleave the current and previous data in the training batches.
Sage could also be prone to \textit{class imbalance}, where the number of traces that meet QoS
is significantly higher than those which violate QoS. In that event, the Data Streamer oversamples the minority class to
incentivize the model 
to generate counterfactuals that violate QoS, and create a more balanced training dataset. 


At runtime, Sage uses the latest version of the GVAE to diagnose QoS violations.
Based on training data, Sage first labels the medians of per-tier performance and usage when QoS is met as \textit{normal values}.
If during execution QoS is violated, the GVAE generates counterfactuals 
by replacing a microservice's performance/usage with their respective \textit{normal values}. 

Sage implements a two-level approach to locate a root cause, to remain lightweight and practical at scale. 
It first uses service-level counterfactuals to locate the culprit microservice that initiated the performance degradation,
and then uses resource-level counterfactuals in the culprit, to identify the underlying reason for the QoS violation and correct it.
More precisely, for each microservice, Sage restores all its metrics 
to their normal values and uses the GVAE to generate the counterfactual end-to-end latency based on the CBN structure.
Since the CBN indicates the causal relationship between a given RPC and the examined microservice, for all non-causally related RPCs,
the GVAE reuses their current per-tier latencies in the counterfactual.
The microservice that reduces the end-to-end latency to just below QoS is signaled as the culprit. After locating the offending microservice, Sage
generates resource-specific counterfactuals to examine the impact of each hardware resource on 
end-to-end performance.
The instantaneous CPU frequency and utilization act as CPU indicators, memory utilization as a memory indicator,
network bandwidth, TCP latency, and ICMP latency as network indicators, etc.
Compared to a one-level approach which tries to jointly locate the service and resource, the two-level scheme is simpler and faster.

Finally, there are cases where multiple microservices are jointly responsible for a QoS violation. 
In such cases, the GVAE iteratively explores microservice combinations when generating counterfactuals,
by adding each time the tier which would have reduced the end-to-end latency the most.



\vspace{-0.04in}
\subsection{Actuation}
\label{sec:design_actuation}
\vspace{-0.04in}

Once Sage determines the root cause of a QoS violation, it takes action. Sage has an actuation controller
in the master and one actuation agent per node. The GVAE notifies the actuation controller, which
locates the nodes with the problematic microservices using service discovery in the container manager, and notifies their respective
actuation agents to intervene. Depending on which resource is identified 
as instigating the QoS violation, the actuation agent will dynamically adjust the CPU frequency, scale up/out the microservice,
limit the number of co-scheduled tasks, partition the last
level cache (LLC) with Intel Cache Allocation Technology (CAT), or partition the network bandwidth
with the Linux traffic control's queueing discipline. The actuation agent first tries to resolve the issue by only adjusting
resources on the offending node, and only when that is insufficient it moves to scale out the problematic microservice on new nodes, or migrate it,
especially for stateful backends, which are almost never migrated.

\vspace{-0.04in}
\subsection{Handling Microservice Updates}
\label{sec:retraining}
\vspace{-0.04in}

A major advantage of microservices is that developers
can easily update existing services or add new ones without impacting the entire service architecture.
Sage's ability to diagnose QoS violations can be impacted by changes to application design and deployment,
such as new, updated, or removed microservices.
Training the complete model from scratch for clusters with hundreds of nodes takes tens of minutes to hours,
and is impractical at runtime. To adapt to frequent microservice changes, Sage instead implements
\textit{selective partial retraining} and \textit{incremental retraining} with a dynamically reshapable GVAE similar to~\cite{yoon2018lifelong},
which piggybacks on the VAE's ability to be decomposed per microservice using the CBN.

On the one hand, with selective partial retraining, we only retrain neurons corresponding to the updated nodes
and their descendents in the CBN, because the causal relationships guarantee that all other nodes are not
affected. On the other hand, with incremental retraining, we initialize the network parameters to those of the previous model,
while adding/removing/reshaping the corresponding networks if microservices are added/dropped/updated.
For example, if a new microservice $B$ is added between existing services $A$ (upstream) and $C$ (downstream), neurons would be introduced
for $B$ in the corresponding networks, and only $A$'s parameters would be retrained.
The combination of these two \textit{transfer learning} approaches allows the model to re-converge faster,
reducing the retraining time by more than $10\times$, especially when there is large fanout in the RPC graph.
To collect sufficient training data quickly after an update, we
temporarily increase the tracing sampling rate until the model converges.


\vspace{-0.08in}
\subsection{Discussion}
\label{sec:discussion}
\vspace{-0.06in}

\noindent{\textbf{Cycles in RPC dependencies: }} Generally, microservice graphs are DAGs, since
cycles between tiers create positive feedback loops, which introduce failures and undermine the
design principles of the microservices model. However, bidirectional streaming RPCs exist between two microservices,
where the client and server both send a message sequence independently within a single request~\cite{grpc}.
This cycle cannot be modeled by the CBN. To eliminate such cyclic dependencies,
we merge both sides of the bidirectional streaming RPC into a metanode with both the client- and
server-side latency, which shares the incoming and outgoing edges of both directions. The GVAE treats the metanode
as a normal microservice. 

\noindent{\textbf{Collecting training data: }} Sage leverages an unsupervised GVAE model that does not require
data labeling. Therefore, it directly uses the tracing data collected in-situ by a cloud's monitoring
infrastructure for training. As with any ML model, the quality of training
data impacts accuracy. 
A primary challenge of cloud performance analysis is handling load variation~\cite{211225}.
Here variation is welcome, as it exposes a more diverse range of behaviors Sage can learn from.
Nevertheless, it is still possible that a well-maintained system with few to no QoS violations has insufficient failure modes to train the model.
In this case, Sage can leverage data obtained through fault injection tests with chaos engineering tools, such as
Chaos Monkey~\cite{basiri2016chaos}, which are already in place in many cloud providers, including
Netflix, Google, and Microsoft~\cite{basiri2016chaos,robbins2012resilience,azurechaos,fbchaos}. 

{\noindent{\textbf{Limitations: }} Sage, as well as other data-driven methods, cannot detect the
source of a performance issue if it has never observed a similar situation in the past. Through the latent variables
in the model, Sage locates the problematic job associated with the root cause and flags it as the issue. 
If the source of the QoS violation is not resource-related, then developers need to be involved to examine if there is a software bug. 

%% file: Methodology.tex
\section{Methodology}
\label{sec:methodology}
\vspace{-0.08in}

\subsection{Cloud Services}
\vspace{-0.08in}

\begin{wrapfigure}[12]{l}{0.25\textwidth}
  \centering
  \vspace{-0.14in}
  \includegraphics[width=0.28\textwidth]{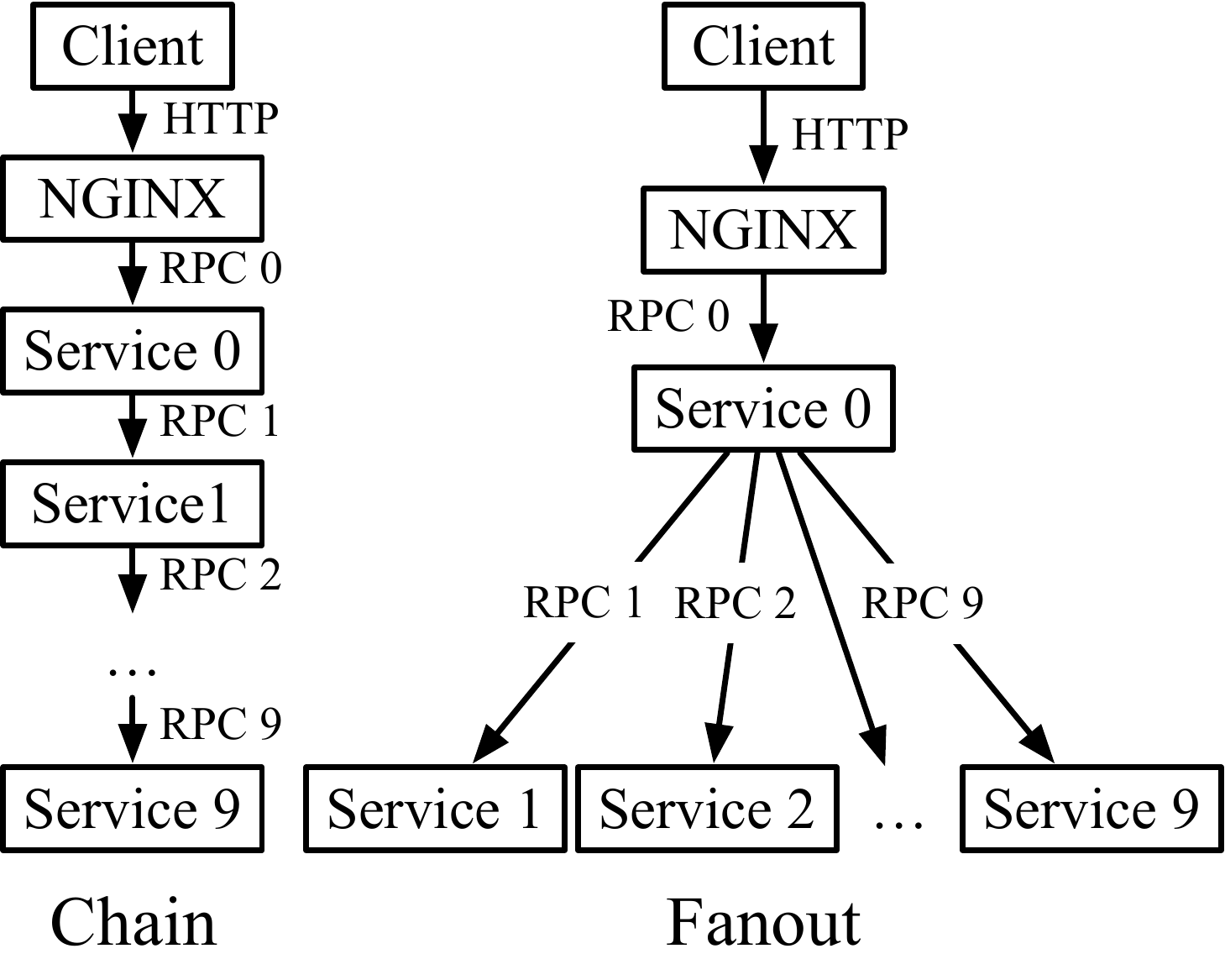}
  \vspace{-0.28in}
  \caption{RPC dependency graph for Chain and Fanout. }
  \label{fig:chain_fanout}
  \vspace{-0.12in}
\end{wrapfigure}

\textbf{Generic Thrift microservices: } \textit{Apache Thrift}~\cite{thrift,slee2007thrift} is a scalable, widely-used RPC framework.
We implement a Thrift code generator to synthesize customizable graphs of resource-intensive
microservices. We can configure the number of microservices, the processing time, the RPC graph, and how RPCs interleave to emulate different functional/timing dependencies.
We generate two common microservice topologies; \textit{Chain} and \textit{Fanout},
shown in Fig.~\ref{fig:chain_fanout}.
\vspace{-0.08in}
In Chain, each microservice receives a request from its upstream service, sends the request to its
downstream tier after processing, and responds to its parent once it gets the results from its child. In Fanout,
the root service broadcasts requests to the leaf tiers, and returns the result to the client only after all children tiers have responded.
We choose the Chain and Fanout topologies because they highlight different behaviors in terms of root cause analysis, and because
most real microservice topologies are combinations of the two~\cite{GrandSLAm,gan:asplos:2019:microservices,usuite}.

\begin{figure}
  \centering
  \includegraphics[width=0.40\textwidth]{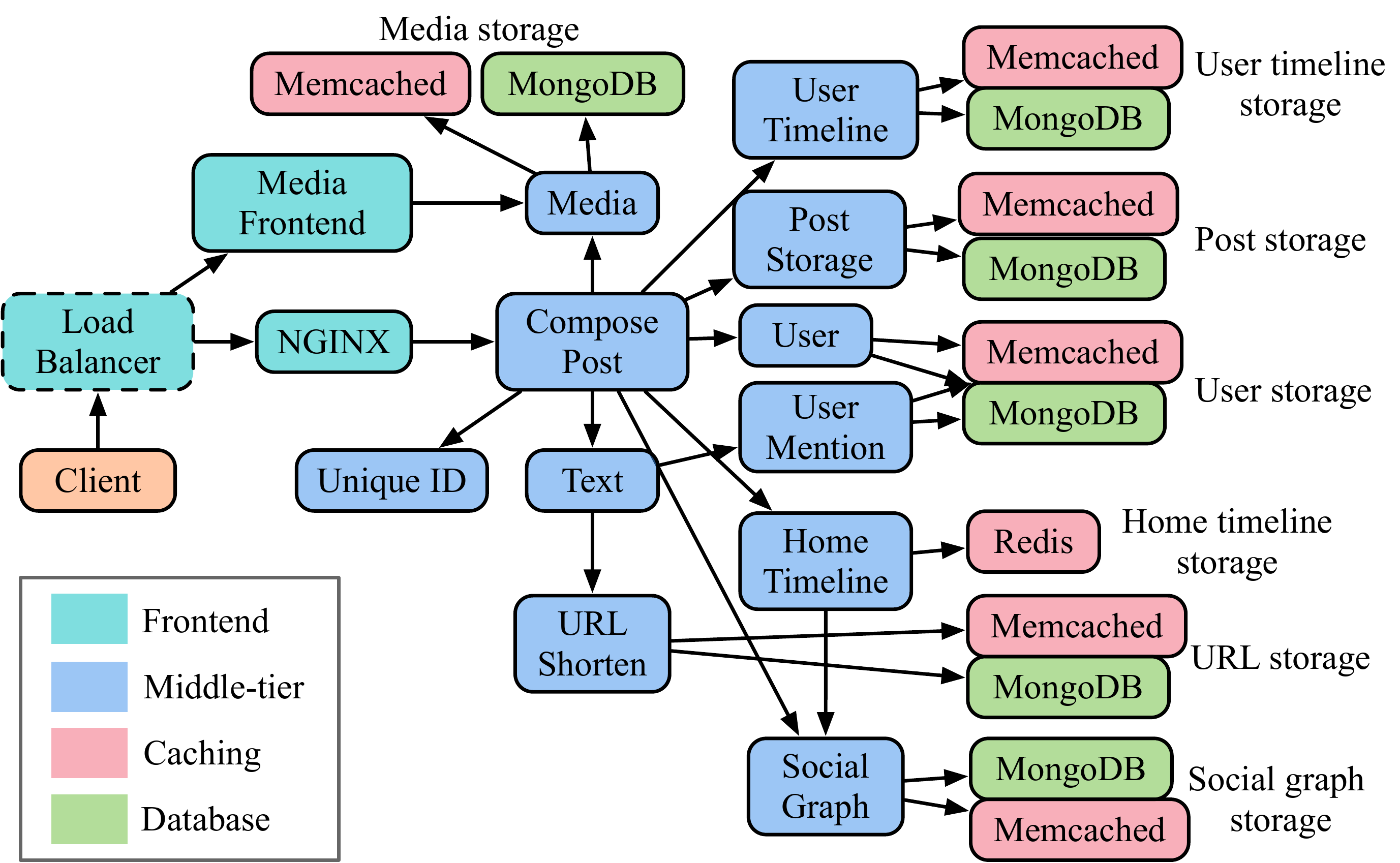}
  \vspace{-0.08in}
  \caption{Social Network microservice architecture~\cite{gan:asplos:2019:microservices}. }
  \label{fig:social_network}
  \vspace{-0.06in}
\end{figure}

\noindent{\textbf{Social Network: }} End-to-end service in DeathStarBench~\cite{gan:asplos:2019:microservices} implementing
a broadcast-style social network. Users can follow/unfollow other users and create posts embedded with text, media, urls, and user mentions,
which are broadcast to their followers. They can also read posts, get user recommendations, and see ads. Fig.~\ref{fig:social_network} shows the Social Network architecture.
The backend uses Memcached and Redis for caching, and MongoDB for persistent storage.
We use the socfb-Reed98 Facebook network dataset~\cite{reed98} as the social 
graph, which contains 962 users and 18.8K follow relationships.

\vspace{-0.08in}
\subsection{Systems}
\vspace{-0.08in}
\noindent{\textbf{Local Cluster: }} We use a dedicated local cluster with five 2-socket 40-core servers
with 128GB RAM each, and two 2-socket 88-core servers with 188GB RAM each. Each server is connected
to a 40Gbps ToR switch over 10Gbe NICs. All services are deployed as Docker containers.

\noindent{\textbf{Google Compute Engine: }} We also deploy the Social Network on a GCE cluster
with 84 nodes in \texttt{us-central1-a} to study Sage's scalability. Each node has 4-64 cores, 4-64GB RAM and
20-128GB SSD, depending on the microservice(s) deployed on it. There is no interference from external jobs. 

\vspace{-0.08in}
\subsection{Training Dataset for Validation}
\vspace{-0.06in}

We use wrk2~\cite{wrk2}, an open-loop HTTP workload generator, to send requests to the web server in
all three applications. To verify the ground truth for Sage's validation in Sec.~\ref{sec:evaluation},
we use \texttt{stress-ng}~\cite{stress-ng} and \texttt{tc-netem}~\cite{tc-netem} to inject CPU-, memory-, disk-,
and network-intensive microbenchmarks to different, randomly-chosen 
microservices, to introduce unpredictable performance. Apart from resource interference, we
also introduce software bugs for Sage to detect, including concurrency bugs and insufficient threads and connections in the pool. 

\vspace{-0.08in}

%% file: Evaluation.tex
\section{Evaluation}
\label{sec:evaluation}

\vspace{-0.08in}
\subsection{Sage Validation}
\label{sec:validation}
\vspace{-0.06in}

\noindent{\textbf{Counterfactual generation accuracy: }} We first validate the GVAE's accuracy in generating counterfactuals
from the recorded latencies in the local cluster. Appropriate counterfactuals should follow the latency distribution
in the training set, but also capture events that are possible, but have not necessarily happened in the past to
ensure a high coverage of the performance space.
There is no overlap between training and testing sets.
We examine the coefficient of determination ($R^2$) and
root-mean-square error (RMSE) of the GVAE in reconstructing latencies in the test dataset.
$R^2$ and RMSE measure a model's goodness-of-fit. The closer to 1 $R^2$ is, and the lower the RMSE,
the more accurate the predictions.
Across all three applications, $R^2$ values are
above $0.91$, and RMSEs are \textcolor{blue}{7.8, 5.1, and 3.2} respectively for the Chain, Fanout and Social Network services,
denoting that the GVAE accurately reproduces the distribution and
magnitude of observed latencies in its counterfactuals. Note that the standard deviations
of latencies in the validation set are high, highlighting that
generating representative counterfactuals is non trivial. 

\begin{figure}
  \centering
  \includegraphics[scale=0.25,viewport=300 30 650 350]{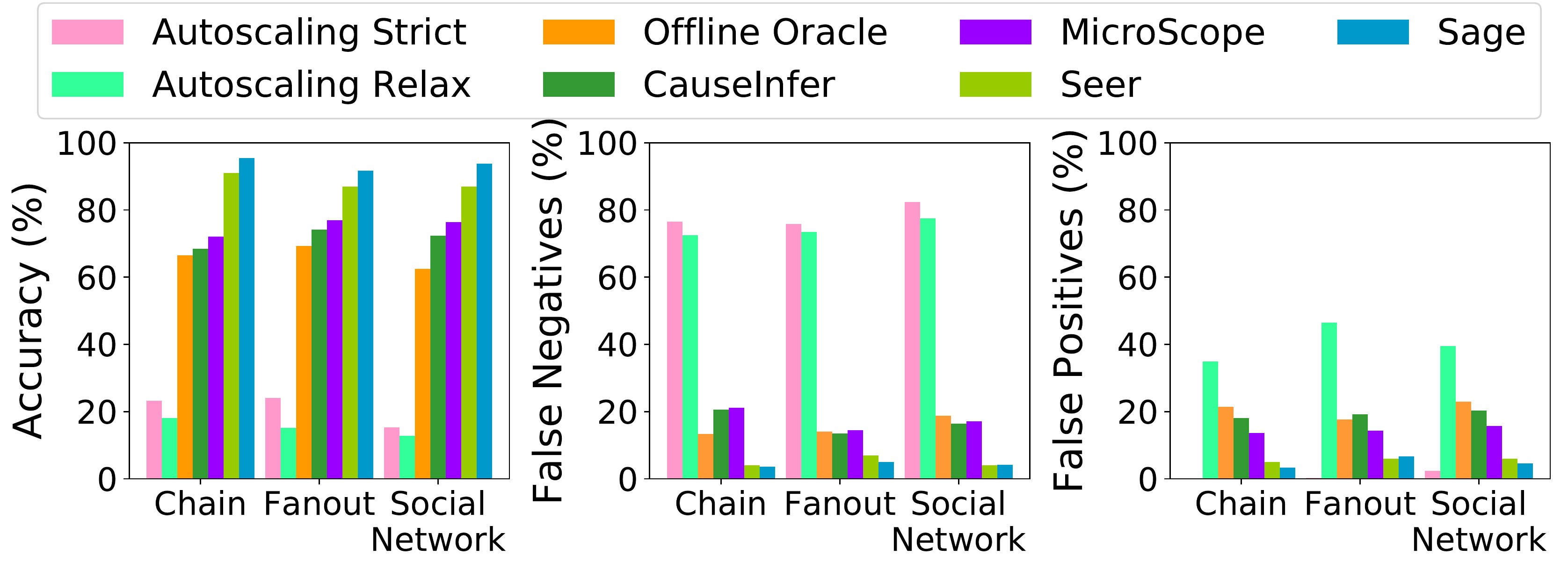}
  \caption{Detection accuracy, and false negatives/positives. }
  \label{fig:accuracy}
  \vspace{-0.06in}
\end{figure}

\noindent{\textbf{Root Cause Diagnosis: }}
Fig.~\ref{fig:accuracy} shows Sage's accuracy in detecting root causes,
compared to two autoscaling techniques, an Oracle that sets upper thresholds
for each tier and metric offline, CauseInfer~\cite{causeinfer}, Microscope~\cite{lin2018microscope}, and Seer~\cite{gan:asplos:2019:seer}. 
\textit{Autoscale Strict} upscales allocations when a tier's CPU utilization 
exceeds 50\%, and \textit{Autoscale Relax} when it exceeds 70\% (on par with AWS's autoscaling policy).
Root causes include both resource-related issues (by injecting contentious kernels in a randomly-selected subset of microservices) 
and software bugs. Since none of the methods do code-level bug inspection, 
a software bug-related issue is counted as correctly-identified if the system identifies the problematic microservice correctly. 

Sage significantly outperforms the two autoscalers and even the offline oracle, by learning
the impact of microservice dependencies, instead of memorizing per-tier/metric thresholds
for a particular cluster state. Similarly, Sage's false negatives
and false positives are marginal. 
False negatives hurt performance, by missing the true source of unpredictable performance,
while false positives 
hurt resource efficiency, by giving more resources to the wrong microservice.
The 3-4\% of false negatives in Sage always correspond to cases where the performance of multiple microservices
was concurrently impacted by independent events, e.g., a network-intensive co-scheduled job impacted one
microservice, while a CPU-intensive task impacted another. While Sage can locate multiple root causes,
that takes longer, and is prone to higher errors than when a single tier is the culprit.
The 3-5\% of false positives are caused by spurious correlations between tiers that were not critical enough
to violate QoS.
Out of the three services, Fanout has slightly lower accuracy, 
due to the fact that a single
misbehaving leaf can significantly impact the end-to-end performance. In general, accuracy varies little between the three services,
showing the generality of Sage across service architectures.

In comparison, the two autoscaling systems misidentify the majority of root causes;
this is primarily because high utilization does not necessarily imply that a tier
is the culprit of unpredictable performance. Especially when using blocking connections, e.g., with HTTP1.1,
bottlenecks in one tier can backpressure its upstream services, increasing their utilization.
Autoscaling misidentifies such highly-used tiers as the culprit, even though the bottleneck is elsewhere.
Additionally, using a global CPU utilization threshold for autoscaling does not work well for microservices,
as their resource needs vary considerably, and even lightly-utilized services can cause performance issues. 
Similarly, the offline Oracle has lower accuracy than Sage,
since it only memorizes per-tier thresholds for a given cluster state, and cannot adapt to changing
circumstances, e.g., load fluctuation, tier changes, or contentious co-scheduled tasks. It can also
not account for tier dependencies, or diversify between backpressure and true resource saturation.

CauseInfer and Microscope have similar accuracy since they both rely on the PC-algorithm~\cite{kalisch2007estimating} to
construct a CPDAG for causal inference. Due to statistical errors and data discretization in computing the conditional
cross entropy needed for the conditional independence test from distributed traces, the CPDAG's structure has inaccuracies, 
resulting in incorrect paths when traversing the graph to identify root causes. 
In contrast, Sage's CBN is directly built from the RPC graph, and considers the usage metrics of different tiers jointly, 
instead of in isolation, 
leading to much higher accuracy. 

Finally, Sage and Seer have comparable accuracy and false negatives/positives; the difference lies in Sage's practicality.
Unlike Seer, which requires expensive and invasive instrumentation to track the queue lengths across the system stack in
each microservice, and additionally relies on supervised trace labeling to learn the QoS violation root causes,
Sage only relies on sparse and non-invasive tracing, already available in most cloud providers. Sage does not require
any changes in the existing application or system stack, and only relies on live
data to learn the root causes of QoS violations, instead of offline training. This makes Sage more practical
and portable at datacenter-scale deployments, especially when the application includes libraries or tiers that cannot be
instrumented. We have verified that Sage is not sensitive to the tracing frequency. 

\begin{wrapfigure}[9]{l}{0.22\textwidth}
  \vspace{-0.16in}
  \centering
  \small
  \begin{tabular}{c||c|c}
    \hline
    Non-         & \multirow{3}{*}{Sage} & \multirow{3}{*}{Seer} \\
    instrumented &                       &                       \\
    tiers        &                       &                       \\
    \hline
    5\%          & 94\%                  & 90\%                  \\
    \hdashline[0.5pt/2.5pt]
    10\%         & 94\%                  & 74\%                  \\
    \hdashline[0.5pt/2.5pt]
    20\%         & 94\%                  & 66\%                  \\
    \hdashline[0.5pt/2.5pt]
    50\%         & 94\%                  & 34\%                  \\
    \hline
  \end{tabular}
  \vspace{-0.06in}
  \caption{\label{missing_info} Accuracy with incomplete instrumentation. }
\end{wrapfigure}
To highlight this, in Table~\ref{missing_info} we show how Seer and Sage's accuracy is impacted from incomplete instrumentation.
For Social Network, we assume that a progressively larger fraction of randomly-selected microservices cannot be instrumented.
Both Sage and Seer can still track the latency, resource usage, - and for Seer, the number of outstanding requests - at the ``borders'' (entry and exit points)
of such microservices, but cannot inject any additional instrumentation points, e.g., to track the queue lengths in the OS, libraries,
or application layer. Even for a small number of non-instrumented microservices, Seer's accuracy drops rapidly,
as queues are misrepresented, and root causes 
cannot be accurately detected. In contract, Sage's accuracy is not impacted, since the system does not require any instrumentation
of a tier's internal implementation.

\vspace{-0.08in}
\subsection{Actuation}
\label{sec:actuation}
\vspace{-0.06in}

\begin{figure}
  \centering
  \includegraphics[scale=0.28,viewport=150 30 700 310]{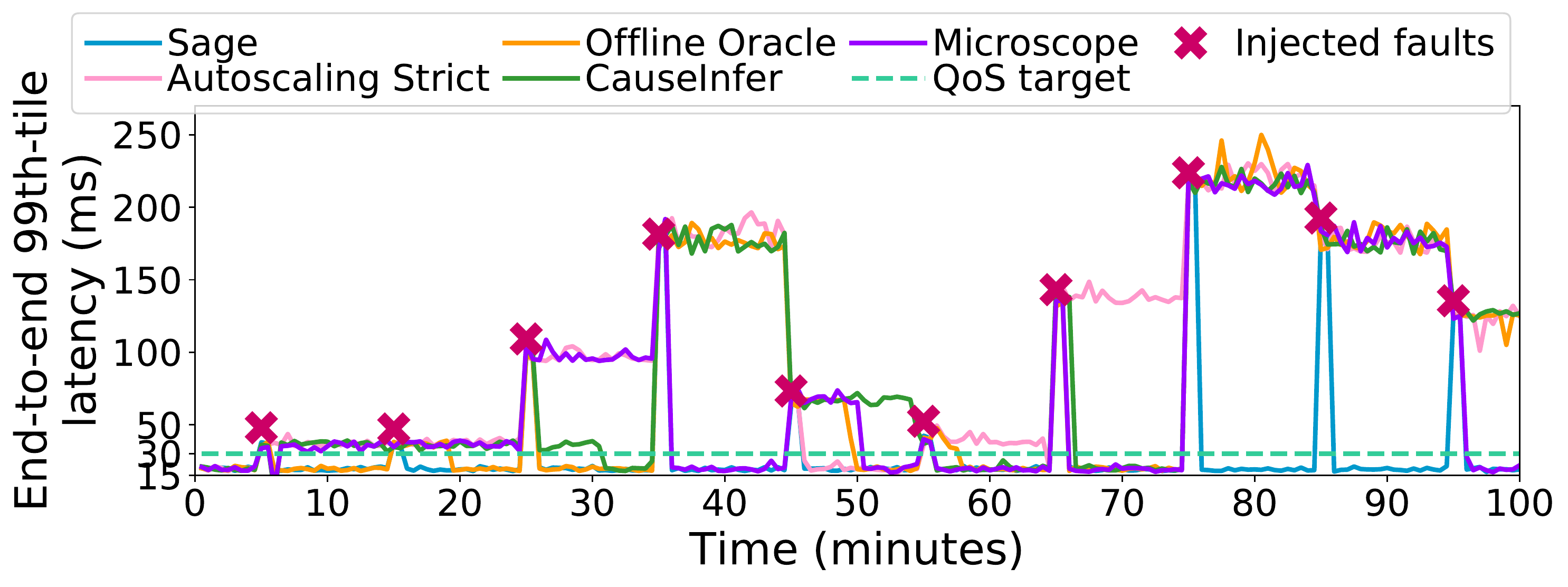}
  \caption{End-to-end tail latency for Social Network. }
  \label{fig:actuation}
  \vspace{-0.08in}
\end{figure}

Fig.~\ref{fig:actuation} shows the tail latency for Social Network managed by Sage,
  the offline Oracle, 
  Autoscale Strict (the best of the two autoscaling schemes), CauseInfer, and Microscope. 
  We run the Social Network for 100 minutes and inject different contentious kernels 
  to multiple randomly-selected microservices. 


Sage identifies all root causes and resources correctly. Upon detection, it notifies the actuation manager
to scale up/out the corresponding resources of problematic microservices.
Inference takes a few tens of milliseconds, and actuation takes tens of milliseconds to several seconds
to apply corrective action, depending on whether the adjustment is local, or requires spinning up new containers. 
In both cases, the process is much faster than the 30-second data sampling interval.
After corrective action is applied the built-up queues start draining; latency always recovers at most
after two sampling intervals from the QoS violation. 
On the other hand, the offline oracle fails to discover the problematic microservices, 
or takes several intervals to locate the root cause,
overprovisioning resources of non-bottlenecked services in the meantime. Recovery here takes
much longer, with tail latency significantly exceeding QoS.
Furthermore, even when the root cause is correctly identified, Oracle often overprovisions
microservices directly adjacent to the culprit, as they likely exceed their thresholds due to backpressure,
leading to resource inefficiency.
The autoscaler only relies on resource utilization, and hence fails to identify
the culprits in the majority of cases, leading to prolonged QoS violations. CauseInfer and Microscope similarly 
do not detect several root causes correctly, due to misidentifying dependencies between tiers, and lead to 
prolonged QoS violations. We omit Seer from the figure as it behaves similarly to Sage.

\vspace{-0.08in}
\subsection{Retraining}
\vspace{-0.06in}

We now examine Sage's real-time detection
accuracy for Social Network, when microservices
are updated.
We roll out six updates, which include adding, updating, and removing microservices from the end-to-end service. 

\begin{figure}
  \includegraphics[scale=0.224,viewport=0 20 700 360]{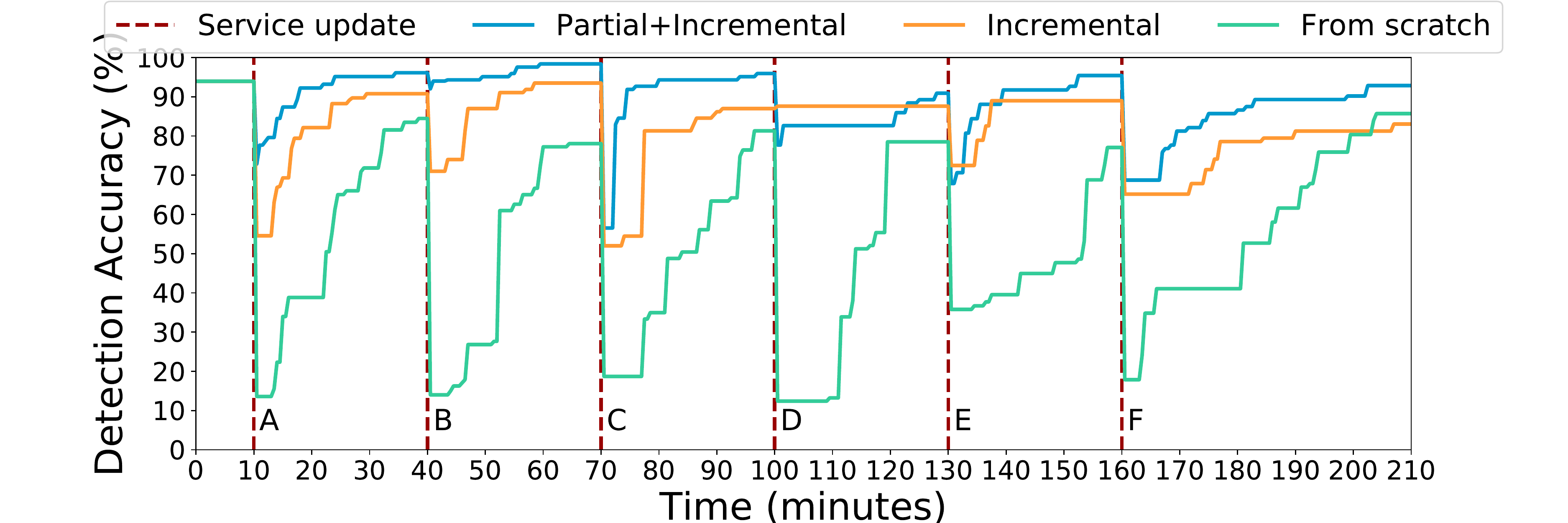}
  \caption{Detection accuracy without/with partial \& incremental retraining.
    Dash lines show when updates are rolled out. }
  \label{fig:incremental_acc}
  \vspace{-0.04in}
\end{figure}

The six updates are indicated by red dash lines labeled with A-F in Figure~\ref{fig:incremental_acc}.
In $A$, we add a new child service to \texttt{compose-post}, close to the front-end, which processes and ranks hashtags.
In $B$, we increase the computation complexity of \texttt{hashtag-service} 
by $5x$. In $C$, we remove the \texttt{hashtag-service}. 
In $D$, we add a new \texttt{url-preprocessing} service closer to the backend, between \texttt{url-shorten}
and \texttt{url-shorten-mongodb}. The further downstream a new service is, the more neurons will have to be updated. 
In $E$, we re-incorporate the \texttt{hashtag-service}, slow down \texttt{url-preprocessing},
and remove \texttt{user-timeline} to capture Sage's behavior under multiple concurrent changes.
In $F$, we revert \texttt{url-preprocessing} and \texttt{hashtag-service} to their previous configurations,
add \texttt{user-timeline}, remove \texttt{home-timeline} and \texttt{home-timeline-redis}, and increase
the CPU and memory requirements of \texttt{compose-post}.


We intentionally create significant
changes in the microservice graph, and 
compare the accuracy of three retraining policies. \textit{Retraining from scratch} creates a new model
every time there is a change, with all network parameters re-initialized.
\textit{Incremental retraining} reuses the network parameters from the previous model, if possible,
and retrains the entire network. \textit{Partial+incremental retraining} uses all techniques of Sec.~\ref{sec:retraining},
which reuse the existing network parameters and only retrain the neurons that are impacted
by the updates.
All approaches are trained in parallel; a new data batch arrives every 30s. 

\noindent{\textbf{Retraining time: }}
Retraining for \textit{partial+incremental retraining}
takes a few seconds and up to a few minutes for the largest data batches. Moreover, it is $3-30\times$ faster
than the other two policies, because it only retrains neurons directly affected by the update,
a much smaller set compared to the entire network. 
The more microservices are updated, and the deeper the updated microservices are located in the RPC dependency graph (updates $D$, $E$, $F$), the higher
the retraining time. 

\noindent{\textbf{Root cause detection accuracy: }}
Fig.~\ref{fig:incremental_acc} shows that \textit{partial+incremental retraining}
and \textit{incremental retraining} have the lowest accuracy drop immediately after an update. 
On the contrary, \textit{retraining from scratch} almost loses its inference ability right after an update,
since the network parameters are completely re-initialized, and the model forgets its prior knowledge. Note that the previous
model cannot be used after the update, because introducing a new microservice changes the GVAE and network dimensions.
\textit{Partial+incremental retraining} converges much faster than the other two
models, because of its shorter retraining time, which prevents neurons irrelevant
to the service update from overfitting to the small training set and forgetting the previously-learned information.

\vspace{-0.08in}
\subsection{Scalability}
\label{sec:scalability}
\vspace{-0.08in}

Finally, we deploy the Social Network on 188 containers on GCE using Docker Swarm.
We replicate all stateless tiers on 2-10 instances, depending on their resource needs, 
and shard the caches and databases. 
We simulate a graph of 1000 users. 

\begin{wrapfigure}[8]{r}{0.28\textwidth}
  \centering
  \includegraphics[scale=0.27, viewport=300 30 300 230]{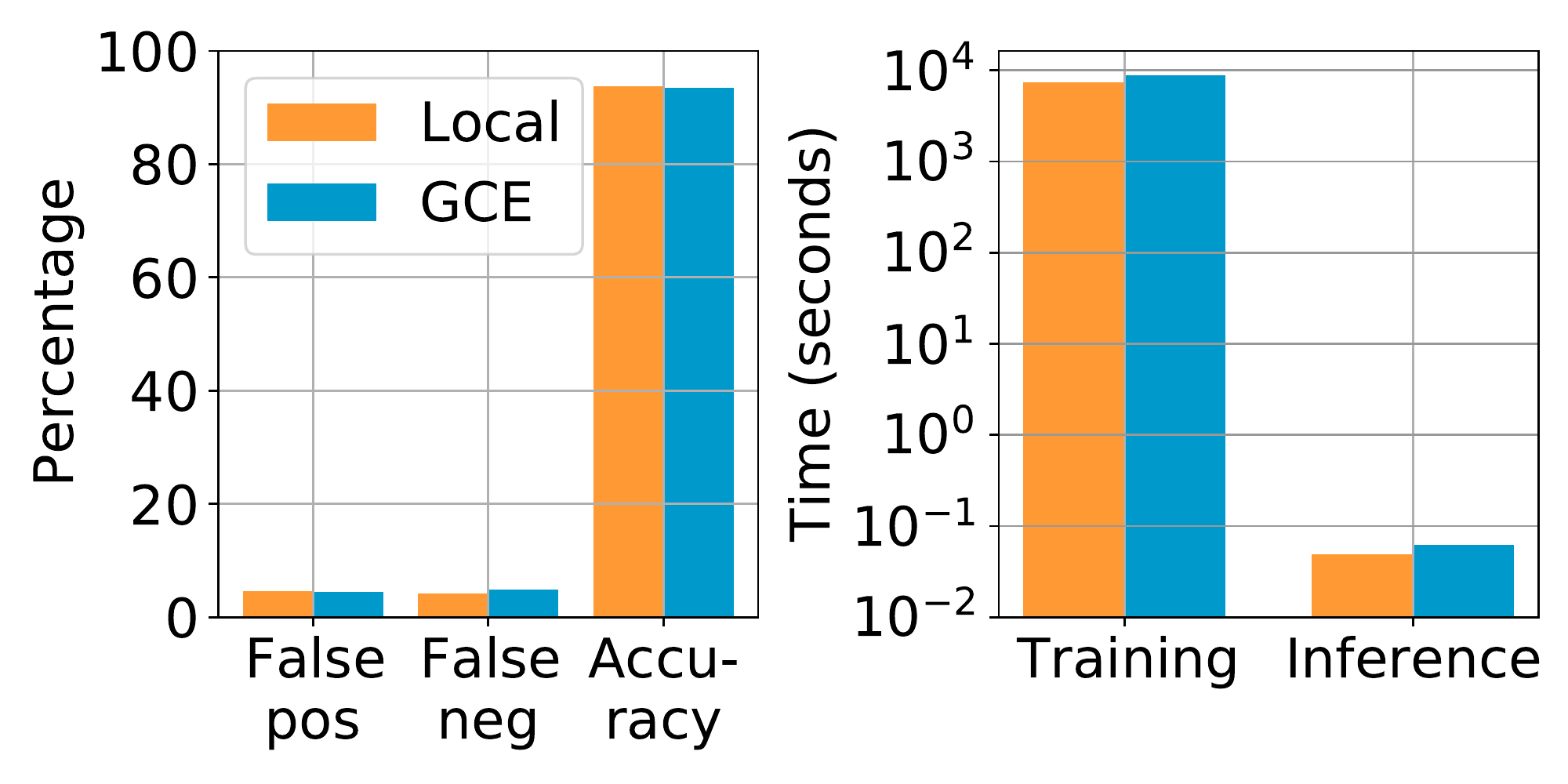}
  \caption{Sage's accuracy and speed on the local cluster and GCE. }
  \label{fig:gce_accuracy}
\end{wrapfigure}

We first validate Sage's accuracy compared to the local
cluster. Fig.~\ref{fig:gce_accuracy}a shows that the accuracy on GCE is unchanged,
indicating that Sage's ability to detect root causes is not impacted by system scale.
Fig.~\ref{fig:gce_accuracy}b compares the training and inference time on the two clusters. 
We use two Intel Xeon 6152 processors with 44 cores for training and inference. Sage takes 124 min
to train from scratch on the local cluster and 148 min on GCE. Root cause inference takes 49ms on the local cluster and 62ms on GCE.
Although we deploy $6.7\times$ more containers on GCE, the training and inference times only
increase by 19.4\% and 26.5\% respectively. In comparison, a similar increase in cluster size, resulted in an almost $4\times$ increase 
in inference time for Seer~\cite{gan:asplos:2019:seer}. 
Sage's good scalability is primarily
due to the system collecting a percentile tensor of latency and usage metrics across all per-tier replicas, 
and due to avoiding high-frequency, detailed tracing for root cause detection. 

\vspace{-0.05in}



%% file: Future_Work_and_Conclusions.tex
\section{Conclusions}
\label{sec:conclusions}

We have presented Sage, an ML-driven root cause analysis system 
for interactive cloud microservices. Unlike prior work, Sage leverages 
entirely unsupervised ML models to detect the source of unpredictable performance, 
removing the need for empirical diagnosis or 
data labeling. Sage works online to detect and correct performance issues, 
while also adapting to changes in application design. In both small- 
and large-scale experiments, Sage achieves high accuracy in pinpointing the root cause 
of QoS violations. 
Given the increasing complexity of cloud services, automated, data-driven systems like Sage improve 
performance without sacrificing resource efficiency.